# A Reduced Complexity Cross-correlation Interference Mitigation Technique on a Real-time Software-defined Radio GPS L1 Receiver

Erick Schmidt
*Department of Electrical and Computer Engineering*
The University of Texas at San Antonio
San Antonio, TX, USA
erickschmidtt@gmail.com

Zach A. Ruble
*Department of Electrical Engineering*
The University of Tennessee at Chattanooga
Chattanooga, TN, USA
zachary-ruble@utc.edu

David Akopian
*Department of Electrical and Computer Engineering*
The University of Texas at San Antonio
San Antonio, TX, USA
david.akopian@utsa.edu

Daniel J. Pack
*Department of Electrical Engineering*
The University of Tennessee at Chattanooga
Chattanooga, TN, USA
daniel-pack@utc.edu

*Abstract*—The U.S. global position system (GPS) is one of the existing global navigation satellite systems (GNSS) that provides position and time information for users in civil, commercial and military backgrounds. Because of its reliance on many applications nowadays, it's crucial for GNSS receivers to have robustness to intentional or unintentional interference. Because most commercial GPS receivers are not flexible, software-defined radio emerged as a promising solution for fast prototyping and research on interference mitigation algorithms. This paper provides a proposed minimum mean-squared error (MMSE) interference mitigation technique which is enhanced for computational feasibility and implemented on a real-time capable GPS L1 SDR receiver. The GPS SDR receiver SW has been optimized for real-time operation on National Instruments' LabVIEW (LV) platform in conjunction with C/C++ dynamic link libraries (DLL) for improved efficiency. Performance results of said algorithm with real signals and injected interference are discussed. The proposed SDR receiver gains in terms of BER curves for several interferers are demonstrated.

## I. Introduction

The U.S. global position system (GPS) is a popular global navigation satellite systems (GNSS) that provides position and time information for users in civil, commercial and military applications. After GPS was made available to commercial and civil markets, the use of such technology rapidly expanded to the point where there is a high dependency on the system for everyday operations. With the arrival of modern technology, such as autonomous cars and pseudolites for indoor navigation, our reliance on GNSS applications is only rising.

With so many existing applications that depend on GPS, it is crucial for the GPS receivers to have robustness in terms of intentional or unintentional interference. GPS signals can experience unintentional interference from other Radio Frequency (RF) bands, or intentional interference due to malicious jamming and/or spoofing attacks. Jamming is the intentional transmission of strong RF signals that overwhelm weaker GPS satellite signal, therefore blinding the target receiver from satellites. On the other hand, spoofing is the clever transmission of counterfeit GNSS-like signals that entail the receiver to compute inaccurate positioning and timing.

GPS interference mitigation techniques is still an active and important topic of research today. There have been many techniques developed to deal with both intentional and unintentional interference over the years, which we separate into two groups: 1) those that use advanced antenna techniques (e.g. arrays) [1]-[5], and 2) those that use IF signal processing (e.g. single antenna) [6]-[11]. Adaptive antenna array methods provide powerful countermeasure solutions which exploit direction-of-arrival (DOA) information. However, this approach typically requires two or more antennas which translates to costly implementations.

Due to the intense computational requirements, conventional GPS receivers use hardware (HW) application specific integrated circuits (ASICs) which provide very efficient computation, but at the cost of limited flexibility and adaptability in implementing emerging GNSS receiver technology. In this context, software defined radio (SDR) seems to be the natural solution, where HW receiver components are replaced with software (SW) that can be reconfigured. The flexibility introduced by using SDR make it an ideal option for fast prototyping and testing of new receiver architectures and algorithms. The major objective of a SW receiver then, is to efficiently implement high rate computations while maintaining a desired amount of flexibility. These two objectives are generally at odds with one another, where an improvement in one feature often occurs at the expense of the other.

Many existing SDR GNSS receivers have been developed with differing levels of HW and SW implementation. These receivers generally use either a combination of field programmable gate arrays (FPGAs) and digital signal processors (DSPs) [12]-[14], a combination of FPGA HW and PC SW [15], or are fully implemented in SW on a PC [16], [17], [18], [19]. Of the mentioned SDR GNSS receivers, only a few of them address interference mitigation. In [20], authors implement and test a real-time SW receiver for detecting the presence of spoofing. Authors in [21] evaluate the performance of interference mitigation using wavelets for Radio Frequency Interference (RFI) and notch filtering for Continuous Wave Interference (CWI) using recorded GPS data on their SW receiver platform *IpexSR* [17]. All of the existing interference mitigation techniques implemented in SDR use either recorded GPS data, or only detect the presence of interference.

This paper provides a proposed minimum mean-squared error (MMSE) interference mitigation technique from [6] which is enhanced for computational feasibility, and implemented on a real-time capable GPS L1 SDR receiver. The GPS SDR receiver SW has been optimized for real-time operation on National Instruments' LabVIEW (LV) platform in conjunction with C/C++ dynamic link libraries (DLL) for improved efficiency as seen in [18], [19].

An interference injection and mitigation testbed was developed to simulate real-time GNSS-like interference with real satellite signals. Simulations show the SDR system was able to reject an interference signal with power levels up to 30dB higher than the satellite signal power. Furthermore, the system was able to reject interference from up to three simultaneous interference signals, each with 30 dB higher power than satellite signal power. Additionally, the proposed receiver is able to inject and correct interference for up to 12 channels in real-time operation mode [18], [19]. The novelty of this approach is the practical application and demonstration of mitigation on GPS-like spoofing signals.

Section II explains GPS L1 overview as well as baseband modules. Section III discusses the proposed algorithm as well as the reduced complexity approach. Section IV explains algorithm implementation in proposed SDR. Section V presents performance result, and Section VI provides conclusion remarks.

## II. GPS L1 BASEBAND MODULES

The flexibility of SDRs showed an increase to numerous software-based GNSS receivers. One of the main contributions of an GNSS SDR receiver is the capability to deal with the large computational complexity calculations in real-time fashion, such as user position, velocity, and time estimation (PVT) [18], [19]. Additionally, interference detection and mitigation algorithms used in conjunction with SDRs largely increase the computational complexity of the receiver. In the following we provide the overall functionality of GPS L1 as divided into three modules: acquisition, tracking, and navigation.

The GPS system consists of 32 satellites orbiting the Earth, which transmit *ranging* signals on the L1, L2, and L5 carrier frequency bands [22]. The payload data contained in each satellite vehicle (SV) transmission is modulated into three layers: a BPSK modulation where payload data is transmitted, a spreading sequence of chips, and a carrier wave which for this proposed SDR is the L1 band at 1575.42 MHz.

GPS utilizes direct sequence spread spectrum (DSSS) modulation which is used to spread slower rate outgoing BPSK navigation data with a higher rate signal. These bits are called chips and the signal itself is called a dispreading code (or ranging code). The ranging code for the C/A L1 signal spreads the signal in bandwidth to creating a noise-resistant spectrum. The ranging code sequence is often called a pseudo-random noise (PRN) code, and consists of a known combination of 1023 chips. The PRN code is transmitted at a rate of 1.023 Mcps (chips per second) and consists of a sequence of known ones and zeros. Therefore, one period of the 1023 chip sequence has a duration of 1 millisecond. The navigation data is BPSK modulated and transmitted at a rate of 50 bps, which is combined with the PRN code, and the L1 carrier wave for passband transmission. Each navigation data bit has a length of 20 ms, therefore there are 20 PRN code periods in each data bit. This redundancy property of navigation bit composition becomes useful for proposed cross-correlation based algorithm.

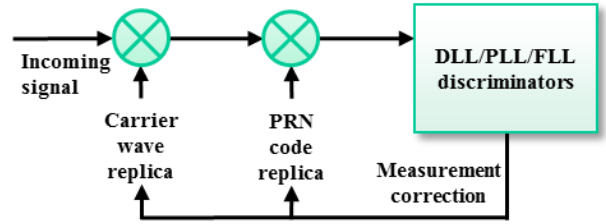

Fig. 1. Conventional tracking feedback loop system for a single channel.

The overall function of a GPS receiver is to continuously synchronize to satellites to obtain navigation data and extract measurements for range estimations. This synchronization is achieved in two steps: (coarse) acquisition to find visible satellite signals and (fine) tracking for regular operation.

Typically, when the SDR receiver front-end (FE) down-converts the incoming signal to baseband or to a known intermediate frequency (IF), residual Doppler Effect carrier sinusoids are still present. Therefore, coarse synchronization (acquisition) for a visible satellite attempts to narrow down a search by attempting correlation matching of both locally generated residual carrier from possible Doppler Effect, and PRN code replicas with the incoming signal to detect an integration peak.

Once a channel has been acquired, a fine synchronization (tracking) is continuously executed. Fig. 1 shows a common tracking feedback loop used on GPS receivers. In a finer synchronization attempt, the tracking module uses closed loop systems to continuously align to the local code and carrier parameters of the visible channel. To determine code and carrier phase measurements of the incoming signal, conventional feedback loops such as delay lock loop (DLL) for code-phase estimation, phase lock loop (PLL) and frequency lock loop (FLL) for carrier-phase estimation, are commonly used [23]. Once these loops output correction measurements, a discriminator processes these measurement outputs to provide filtered quantities which adjust current channel tracking parameters for next (epoch) iteration.

For navigation, once successful residual carrier wipe-off and code dispreading occurs, BPSK data previously modulated at 50 Hz is extracted by integrating 20 ms navigation bits. Navigation data contains crucial information such as satellite positions, and timestamps from the ranging signals. Ranging measurements extracted from tracking loops as well as navigation data is used to execute PVT solutions via trilateration techniques, which are detailed further in [23].

Regarding PRN codes, each of 32 satellites transmits its own unique code. The PRN codes, also called gold codes, have special properties of orthogonality and cross-correlation, thus are barely interfering among each other. The navigation data which is transmitted by each SV spread with its respective PRN code, contains required information for the position solution. The proposed algorithm particularly enhances the previously detailed tracking loop in the presence of interference and will be detailed in next sections.

As for the proposed SDR GPS L1 receiver, it is LV-based and has real-time capabilities for up to 12 tracking channels. For further details on the implementation of proposed SDR platform specifically on SW and HW architectures as well as its configuration options, the reader is directed to [18], [19].

## III. Proposed Algorithm

The method proposed in this paper mitigates the interference phenomena through an optimization approach and by integrating interference mitigation filters onto tracking loop correlators. Two tasks are addressed: (a) tuning the proposed signal processing approach in [6] for stable operation; (b) implementing a real-time version of the algorithm on previously presented SDR platform with variable-length real signals and attainable computational complexity.

For the first task, previously suggested interference mitigation concept in [6] is adapted with a reduced complexity approach to a real environment. As mentioned in Section II, dispreading code replicas are periodically aligned and correlated with fragments of signals transmitted by GPS satellites. These codes are used for both separation (filtering) of target satellite signals for payload data extraction, and for continuous synchronization of satellite signals with locally generated replica codes. The proposed idea of interference mitigation *enhances* the receiver dispreading code for two concurrent reasons: to match and extract target signal (as a typical GPS operation), and for interference signal filtering.

The effect of cross-correlation interference can occur by multiple access interference (MAI) commonly found in spread spectrum (SS) systems [6]. The ensemble of transmitted PRN codes on a composite SS signal can increase interference levels based on transmitter power and relative distance to receiver, therefore disturbing local dispreading code synchronization with received signals. While conventional dispreading codes are fixed and known, the proposed method exploits adaptive modifications of these codes to minimize interference caused by cross-correlating interference. The method uses modifications of the local replica dispreading codes to serve concurrently as synchronization correlators, and interference filters. The corrections of the conventional dispreading code used in the correlator serve as filter components and performs a task that is similar to blind equalization with interference cancellation. To find the most favorable code corrections, an optimization problem is formulated for the minimization of interference using a mean squared error (MSE) cost function. Fig. 2 shows an example of modified dispreading code vs. conventional for an optimal interference filtering.

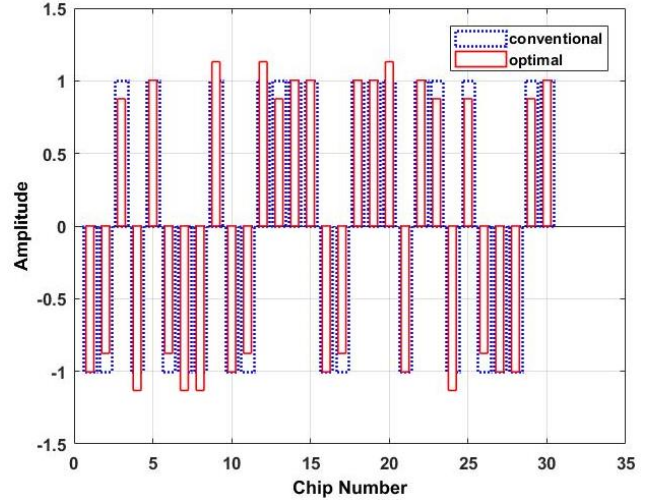

Fig. 2. Optimal dispreading code modifications example for MAI cross-correlation mitigation [6].

The proposed methods is based on the convex optimization theory where the MSE is regarded as a convex function in the dispreading code, as in [6]. At the optimal point (optimal dispreading code), the algorithm maximizes the signal-to-interference-plus-noise ratio (SINR) for the weak satellite signal.

### A. Problem formulation

For the problem formulation, the following is defined. A GPS receiver is assumed to be synchronized to a channel $k$ and residual carrier is negligible (wiped-off). Therefore, the received signal is a fragment of a navigation bit aligned in code and carrier phase, but with carrier phase wipe-off only, thus still containing the spreading code from the current channel. A received signal vector, $\mathbf{r}_k$, for channel $k$ is a superimposition of the currently aligned dispreading code, $\mathbf{s}_k^0$, along with other satellite codes (denoted as interferences), and noise. This received vector $\mathbf{r}_k$ is also associated with its corresponding signal power, $p_k$, and a modulated bit-period polarity, $b_k$, as seen in [6]. This received signal vector can be written as follows:

$$\mathbf{r}_k = b_k \sqrt{p_k} \mathbf{s}_k^0 + \mathbf{i}_k + \mathbf{n}_k \qquad (1)$$

where $\mathbf{i}_k$ is the interference of other satellites, and $\mathbf{n}_k$ is the noise. The received signal is then code wiped-off by multiplying to a locally generated dispreading code, $\mathbf{h}_k$, and integrating the result. The integrated result, or decision variable, $d_k$, is a correlator associating received signal, $\mathbf{r}_k$, and dispreading replica, $\mathbf{h}_k$, for the channel $k$ as follows:

$$d_k = \mathbf{h}_k^T \mathbf{r}_k \qquad (2)$$

where $\mathbf{h}_k$ is a unit norm vector that does not amplify or attenuate the received power during a bit-period. As seen in [6],

each receiver channel $k$ defines and minimizes a mean-squared error (MSE) cost function denoted as $MSE_k$. The MSE cost function for each satellite $k$ can be written as follows:

$$MSE_k = E\left[\left(b_k - \frac{d_k}{\sqrt{p_k}}\right)^2\right] \quad (3)$$

where $E[\cdot]$ is the expected value, and $p_k$ is the received power used to normalize the correlation value $d_k$. The cost function is reduced by a minimization of a quadratic function and linearization by certain conditions. The final solution is known as a solution of a linear equation in [6]:

$$\mathbf{h}_k = p_k \mathbf{R}_k^{-1} \mathbf{s}_k^0 \quad (4)$$

where $\mathbf{R}_k^{-1}$ is matrix inverse if it exists, or pseudoinverse for a minimum norm solution. A consideration at hand is the implementation feasibility of the minimum MSE (MMSE) receiver in (3). The computational cost of a large size matrix inversion $\mathbf{R}_k^{-1}$ as seen in (4), is typically very high with a complexity order of $O(N^3)$, where $N = 1023$ chips.

### B. Reduced complexity implementation

A reduced complexity *group-weighting* method that trades off complexity vs performance in the code adaptation is proposed. This method decreases the computation complexity by collapsing the original dimension of the optimization space (dispreading code size), $N$, to another dimension, $M$, with $M < N$. The procedure is based on the definition of $M$ new orthogonal dimensions by grouping the elements of the received signal vector, specifically chips. Without the loss of generality, we oversample the received signal such that the new space dimension is $N = 1024$ (size power-of-two). This new dimensional space $M$ can be seen as a group of partial correlations with size $g$. Fig. 3 shows the interaction of these partial correlations with input samples and how they estimate corrections for a sub-optimal solution in $d_k$. Each partial correlation is used to build an autocorrelation matrix to obtain an optimal solution similar to (4), where the coefficients maximize the SINR in the MMSE receiver, as explained next.

For the *group-weighting* method proposed in [6], we restrict the dispreading sequence to the following format:

$$\mathbf{h}_k = \mathbf{w}_k (.*) \mathbf{s}_k^0 \quad (5)$$

$$\mathbf{w}_k = \left[(w_1,\ldots,w_1),(w_2,\ldots,w_2),\ldots,(w_M,\ldots,w_M)\right]^T \quad (6)$$

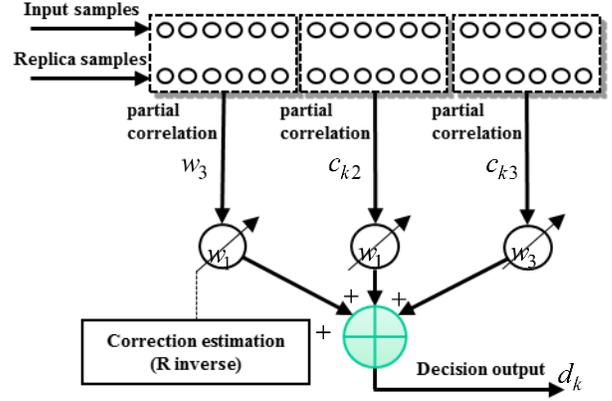

Fig 3. Group-weighting method showing sample interactions with partial correlations.

where vector size $N = g \cdot M$. As the elements of $\mathbf{s}_k^0$ are $\pm 1$, then $\mathbf{h}_k^T \mathbf{s}_k^0 = g \mathbf{w}_k^T \mathbf{1}_M$. Let us split $\mathbf{h}_k$, $\mathbf{r}_k$ and $\mathbf{s}_k^0$ into $M$ segments:

$$\mathbf{h}_k = \begin{bmatrix} \mathbf{h}_{k1} \\ \mathbf{h}_{k2} \\ \vdots \\ \mathbf{h}_{kM} \end{bmatrix}; \mathbf{r}_k = \begin{bmatrix} \mathbf{r}_{k1} \\ \mathbf{r}_{k2} \\ \vdots \\ \mathbf{r}_{kM} \end{bmatrix}; \mathbf{s}_k^0 = \begin{bmatrix} \mathbf{s}_{k1}^0 \\ \mathbf{s}_{k2}^0 \\ \vdots \\ \mathbf{s}_{kM}^0 \end{bmatrix} \quad (7)$$

The constrained $\mathbf{h}_k$ will be then as follows:

$$\mathbf{h}_k = \begin{bmatrix} \mathbf{h}_{k1} \\ \mathbf{h}_{k2} \\ \vdots \\ \mathbf{h}_{kM} \end{bmatrix} = \begin{bmatrix} w_1 \mathbf{s}_{k1}^0 \\ w_2 \mathbf{s}_{k2}^0 \\ \vdots \\ w_M \mathbf{s}_{kM}^0 \end{bmatrix} \quad (8)$$

Denote $c_{kj} = \mathbf{h}_{kj}^T \mathbf{r}_{kj}$ as a partial correlation, and $\mathbf{c}_k = [c_{k1}, c_{k2}, \ldots, c_{kM}]^T$ as a vector of partial correlations. Then,

$$d_k = \mathbf{h}_k^T \mathbf{r}_k = \mathbf{w}_k^T \mathbf{c}_k \quad (9)$$

With proposed substitutions in (7) and (8), we can then proceed similarly as in (3) to obtain the cost function solution for the reduced complexity *group-weighting* method as follows:

$$\mathbf{w}_k = g p_k \mathbf{R}_{ck}^{-1} \mathbf{1}_M \quad (10)$$

The solution in (10) is sub-optimal for $g > 1$ in comparison with the one provided by the optimal algorithm in (4). This is a result of the *group-weighting* method decreasing the freedom of the modified dispreading code, but on the other hand, it has the advantage of significantly reducing the computational complexity (see Fig. 3). Moreover, the solution presented for partial correlations can be implemented with a computational complexity of $O(M^3)$ and a computational gain of $O(g^3)$.

## IV. Algorithm Implementation on SDR Testbed

This section details the MMSE algorithm implementation in real-time execution as a modification to conventional GPS tracking loops from SDR receiver in [18], [19]. An important requirement for the MMSE algorithm to properly detect and correct interference is the autocorrelation matrix which is obtained from the correlators. Autocorrelation matrix obtained from the outputs of the partial correlators serves as a "blind sensing" unit of the interference. It is used for the weight adjustments to filter interference as in (10). Some options to control computational complexity of these updates are the grouping parameter $g$, and window size $L$, which are detailed next.

Fig. 4 shows the MMSE correlator integrated with the tracking loops as well as the autocorrelation matrix $\mathbf{R}_{ck}$ of partial correlations for group-weighting method. Conventional tracking loops have variable sample integration lengths because of Doppler Effects. In the proposed SDR, the tracking loop enforces a constant length of 1023 chips, using a pre-integration within one chip (i.e. one may have 4-5 samples per chip for a sampling rate of 5 MHz). Initially, the tracking loop collects a fixed block size of data which corresponds to 1 ms (~5000 samples for 5 MHz sampling rate). The local carrier replica is sampled as the size of the current epoch (1 ms) considering Doppler Effects (~5000 samples). After carrier wipe-off, pre-integrated within each chip occurs, according to Doppler Effect and residual carrier and code parameters. An array of $N = 1023$ chips is now available. For the *group-weighting* method, the array is upsampled to $N = 1024$ (i.e. powers-of-two) for proper partial correlation grouping. Fig. 4 details the tracking loop modifications beginning after the carrier wipe-off.

After chip pre-integration and chip upsampling, received vector, $\mathbf{r}_k$, is mixed with aligned in-prompt dispreading code replica, $\mathbf{s}_k^0$, for code wipe-off. After this, the *partial integrate and dump* block integrates sequence of chips into $M$ partial correlation groups, outputting $\mathbf{c}_k$ vector of size $1 \times M$ as seen in Fig. 4. Strictly speaking, and for reduced computational complexity, the *partial integrate and dump* block performs both chip pre-integration and grouping in one single step.

The *recursive autocorrelation computation* block obtains $\mathbf{c}_k$ vector as input for $\mathbf{R}_{ck}$ autocorrelation matrix estimation. Once the autocorrelation matrix has collected samples with apposite statistical significance, the MMSE solution is calculated by using (10) for *group-weighting* solution. This is done at the *interference mitigation algorithm (MMSE)* block seen in Fig. 4 to obtain sub-optimal $\mathbf{w}_k$ coefficients vector. Subsequently, this solution vector of size $M$ is correlated to current partial correlation vector $\mathbf{c}_k$ for corrections. Other inputs to the MMSE block are group size, $g$, and channel $k$ signal power, $p_k$, the latter which is collected from tracking loops.

Finally, the *complete integrate and dump* block is used to integrate the result $\mathbf{w}_k \cdot \mathbf{c}_k$ for filtered bit-sample epoch extraction. As a summary, three integration steps are used in the modified tracking MMSE correlator loop, by which the first two are done in one step at the *partial integrate and dump* block, as seen in Fig. 4: chip pre-integration (from samples to chips, i.e. $5000 \rightarrow 1023$), partial integration (from $N$ to $M$, based on $g$ parameter), and complete integration (from $M$ to 1, for navigation bit sample extraction). When comparing modifications to conventional GPS correlator as in Fig. 4, the MMSE implementation can be seen as a block replacing part of a typical integrate and dump (ID) filter.

The implementation of the *recursive autocorrelation computation* $\mathbf{R}_{ck}$ block seen in Fig. 4 is achieved by a recursive statistical method with attainable computational complexity for the current SDR application. Consider a sequence of partial correlations $\mathbf{c}_k(t_l), l = 1,...,t_L$, that is used to estimate autocorrelation matrix $\mathbf{R}_{c_k}$ of vectors $\mathbf{c}_k(t_l)$, such as:

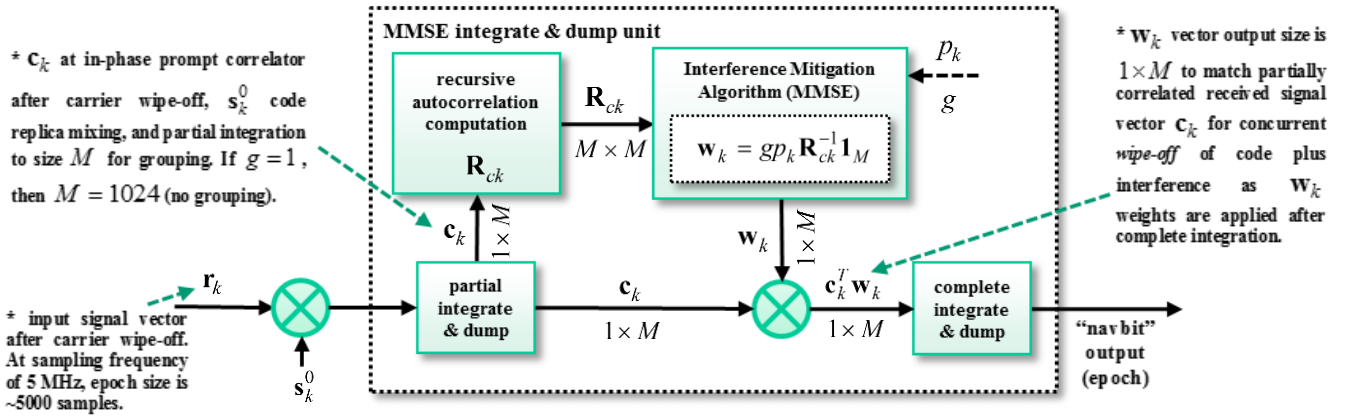

Fig. 4. MMSE correlator implementation in SDR tracking loop after carrier wipe coming from the in-phase prompt channel.

$$\mathbf{R}_{ck}(t_L) = \frac{1}{L}\sum_{l=1}^{L} \mathbf{c}_k(t_l)\mathbf{c}_k(t_l)^T \qquad (11)$$

where $t_l$ is the time instant of $\mathbf{c}_k(t_l)$ availability and $L$ is the window size fixed at time of system initialization. To address complexity constraints for proposed SDR implementation, the autocorrelation matrix is computed recursively in a sliding window manner by discarding the oldest $\mathbf{c}_k(t_l)$ entry and adding the newest one:

$$\begin{aligned}\mathbf{R}_{ck}(t_L+1) &= \mathbf{R}_{ck}(t_L) \\ &\quad - \frac{1}{L}\mathbf{c}_k(t_1)\mathbf{c}_k(t_1)^T \\ &\quad + \frac{1}{L}\mathbf{c}_k(t_L+1)\mathbf{c}_k(t_L+1)^T\end{aligned} \qquad (12)$$

This recursively computed matrix collects several partially correlated vector epochs $\mathbf{c}_k$ of length $M$ (based on the $g$ parameter for partial correlations) on a first-input first-output (FIFO) vector buffer. Therefore, on every epoch the oldest vector in the buffer is subtracted from $\mathbf{R}_{ck}$, and the newest vector is added. These operations on the recursive accumulator occur on each epoch iteration as seen in (12).

For C/C++ implementation, both the size of the FIFO vector buffer as well as the matrix $\mathbf{R}_{ck}$ have to do with the grouping parameter $g$, the vector length $M$, and the sliding window size $L$. Eventually, the total size of the sliding window array is of size $M \times L$ samples and the matrix of size $M \times M$ samples, respectively. The *FIFO vector buffer* block and matrix $\mathbf{R}_{ck}$ are implemented in the C/C++ code by using dynamic allocation based on initial user parameters, $g$, and $L$, prior to run-time execution. Once these sizes are set and initialized, they are used throughout the execution of the SDR system. There are several recommendations for the grouping size $g$ and sliding window $L$ pairs: $g=1, L=1500$ ; $g=2, L=1000$ ; $g=4, L=1000$ ; $g=8, L=1000$ ; $g=16, L=500$ ; $g=32, L=600$ ; $g=64, L=300$. The mostly used grouping size configuration pair which achieves best trade-off between computation complexity and performance is $g=64$, and $L=300$, which allocates a dynamic variable of size 4,800 samples ($g=64$ gives $M=16$).

### A. Interference Injection

Interference injection consists of replicating current channel $k$ aligned code replica, $\mathbf{s}_k^0$, and delaying it in number of chips (this is on top of real signal), i.e. $\mathbf{s}_k^\alpha$, where $\alpha$ is a delay in chips based on known cross correlation properties of the pseudonoise sequence for satellite $k$, thus *spoofing* the real navigation bits which contain navigation data required for positioning. Fig. 5 shows an example of cross correlation of a known gold sequence for satellite vehicle (SV) 1 with itself, omitting the center value at zero (which values is $N$) for scaling purposes.

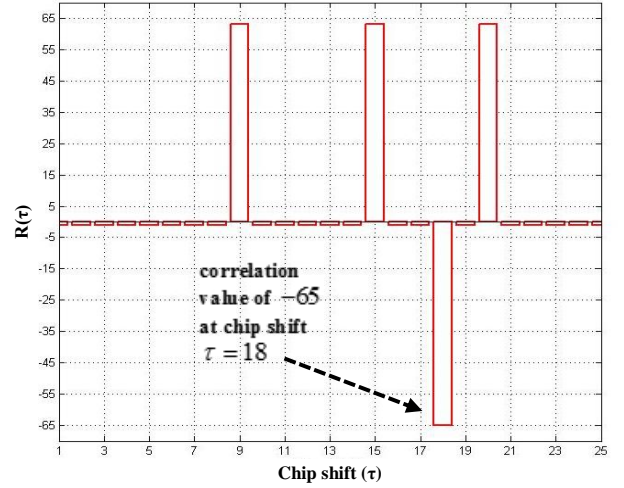

Fig. 5. Gold sequence autocorrelation for SV1 (from shift value 1).

Typically, gold sequences are selected by using m-sequence preferred pairs with good cross correlation properties [24], but still have *disagreement shifts,* in chips, where correlation value is highest between any sequence, thus losing orthogonality. This highest correlation value between any two gold sequences is: $2^{\frac{n+2}{2}}+1$, where $n$ is the degree of primitive polynomial used in preferred pair m-sequences. In GPS systems, $n=10$, therefore the maximum correlation value for any two gold sequences is $65$. In Fig. 5, for SV1, a delay of $\tau=18$ chips will highly enhance MAI with respect to synchronized channel $k$.

In addition, random $\pm 1$ navigation sample bits are modulated onto the delayed interference replica version of the current channel $k$, assigning random bit polarity every 20 samples (20 milliseconds). The received navigation bit and spoofed navigation bit are not synchronized in terms of epochs, since the interference navigation bits can be injected at any time. The worst-case scenario is when both navigation bits are fully synchronized. Power is also typically higher on the injected interference, and is measured in dB of relative higher power with respect to current channel power $p_k$. The resulting signal that is injected before the code wipe-off can be modelled as $\mathbf{i}_k = p_{(i)k} b_{(i)k} s_{(i)k}^\alpha$.

There are 6 correlators in total in a conventional GPS tracking loop for each channel: in-phase early, prompt and late; and quadrature early, prompt and late. The correlator that is used to extract the navigation bits is the in-phase prompt channel, which is where payload data coming from satellites is modulated [23]. All these six channels are used for feedback into the synchronization loops, namely PLL/DLL/FLL corrections for each next epoch iteration. If testing only on navigation data cleaning capabilities with MMSE solution, then a separate 7[th] correlator channel is created to bypass effects on synchronization.

Fig. 6 shows how conventional GPS tracking loop system remains unmodified by adding a 7th tracking arm serving as the *MMSE integrate and dump* block (see Fig. 4) at the top part which is directly obtained from the in-phase prompt arm. This allows to successfully bypass synchronization loop operation and concentrate on data manipulations. The red dot along this extra correlator arm is where the interference is injected, with a known *c* chip delay which follows the prompt code replica delay of that channel, and *the MMSE integrate and dump* block is where corrections occur to obtain clean navigational data. The navigational data is now extracted from this 7th arm, as opposed to conventional GPS tracking loop. This allows a successful isolated simulation environment of data corruption and correction without affecting synchronization loops. With the use of real data input and assuming normal working conditions on the synchronization loops, the system can successfully test the spoofing attack and correction capabilities only on the navigational data.

The interference that is simulated in the SDR testbed is considered a worst-case scenario since it is perfectly aligned in chip-level with local code replica, and it assumes carrier alignment as well. This creates the most damage to target channel as interference injection and therefore can be used as a base. This approach makes the receiver immune to GPS-like synchronized intervention.

*1) On-the-fly configuration*

On-The-Fly (OTF) configuration found on the SDR front panel allows for interference injection and MMSE corrections to be configured in real-time for several channels. The interference injection is added directly to the baseband signal, with respect to the current channel $k$ power $p_k$. It is configured in dB scale, i.e. if 10 dB is set for interference injection, then the delayed interference signal power will be 10 times stronger than current signal power; therefore, interference signal power follows current signal power, which is updated in real-time. Internally, an interference delay table previously generated is stored so that the SDR platform is capable of injecting up to three interferers per channel $k$, namely $\mathbf{s}_k^{\alpha_1}, \mathbf{s}_k^{\alpha_2}, \mathbf{s}_k^{\alpha_3}$, for worst-case chip delay $\alpha$ for current tracking channel. The SDR allows up to 12 channels to run concurrently with MMSE and interference injection algorithms, each with independent power in dB relative to the current channel received power.

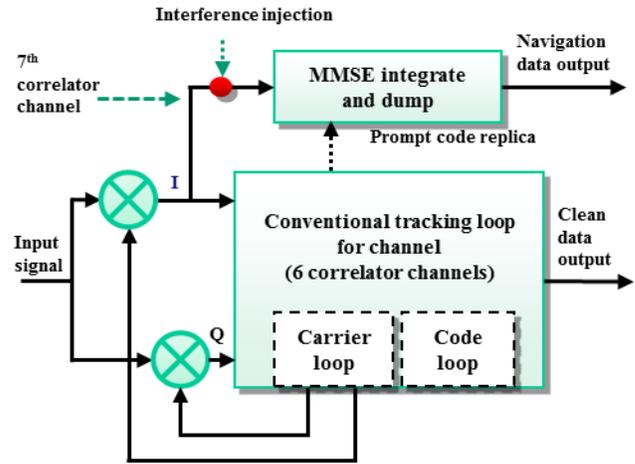

Fig. 6. Interference injection and correction on a 7th correlator channel, bypassing synchronization loops.

Fig. 7 shows OTF configuration front panel (left) which can be manipulated in real-time operation, as well as Channel Health tab (right) visualization panel. The Channel Health tab displays three relevant LEDS: *MMSE* correction (blue), *Interferer* injection (red), and *ValidPVT* navigation solution (green). These three LEDs provide details on whether the current channel is being: corrected with MMSE algorithms, interfered with based on OTF configuration, and if the current channel is being used for PVT solution. The *ValidPVT* LED is useful in showing when interference power is strong enough for the receiver to disregard said channel.

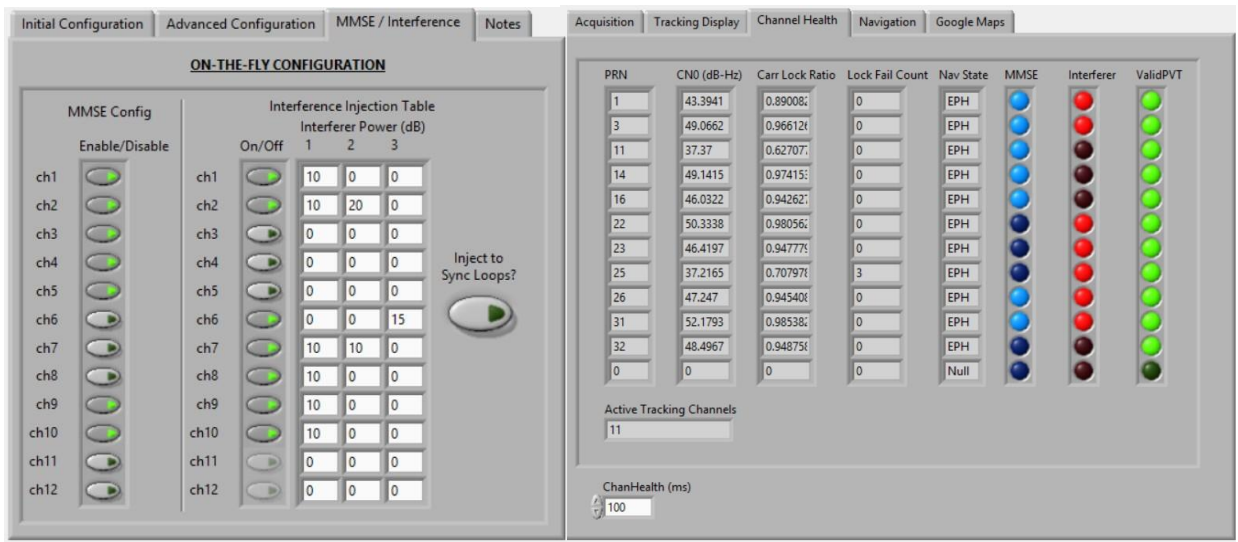

Fig. 7. On-the-fly configuration front panel for proposed SDR with 12 channel configuration capabilities in real-time. Configuration for both MMSE and interference injection is seen on left side panel. Channel health and other status seen on right side panel.

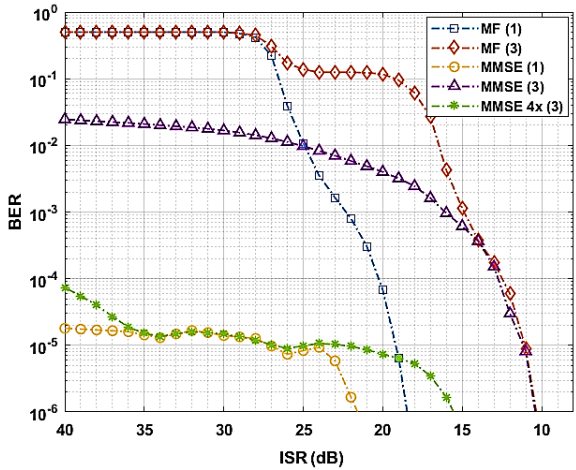

Fig. 8. BER vs SIR-1 performance results for one and three interferers for matched filter and proposed MMSE approach.

TABLE I.  RECEIVER GAIN FOR TARGET BER = $10^{-6}$

| BER = $10^{-6}$ | | |
|---|---|---|
| **Grouping parameter and window size** | *g=64, l=300* | *g=64, l=1200* |
| *1 interferer MF vs. MMSE gain (dB)* | 2.1 dB | 7.3 dB |
| *3 interferers MF vs. MMSE gain (dB)* | 0 dB | 5.2 dB |

Fig. 8 shows performance curve for one and three interferers where MMSE has been assessed for 20 epoch integration period BER. When comparing 1 interferer on MF vs MMSE correlator, a gain of 2.1 dB is observed, with parameters $g=64$ and $l=300$. When increasing the window size four times, a gain of 7.3 dB is observed for one interferer. This same comparison for 3 interferers reflects no gain at same BER, but when MMSE window size is increased four times, an improvement of 5.2 dB is observed. Table I summarizes these BER gain results.

## V. SIMULATION RESULTS

The proposed SDR platform and interference mitigation technique is analyzed by using a hybrid combination of real signals coming from a LabVIEW-based NI GPS simulation toolkit [25] and internal interference that generates current satellite pseudocodes with a delay to cause cross-correlation jamming (see Section IV-A). The SDR SW is able to successfully jam up to 12 channels so that the receiver is not able to obtain necessary payload data from satellites for PVT solution. Moreover, the receiver is able to remain locked with visible satellites during operation since interference bypasses tracking loop synchronization. We test the algorithm with the reduced complexity approach grouping parameter $g=64$ and window size of $l=300$. As an aggregate experiment, we test also $g=64$ and $l=1200$, which is four times the window size from previous parameters. We test these scenarios with 1 and 3 interferers and we compare against a conventional matched filter (MF) correlator.

Without the loss of generality, we evaluate system performance by simulating interference power relative to signal power, in dB scale. We use interference-to-signal (ISR) ratio as noise power can be neglected when compared to total interfering power and signal power after correlation and signal integration. We then plot bit-error rate (BER) vs. ISR performance curves. The interference that is injected is aligned within chips and within navigation bit alignment, thus assuming worst-case scenario for jamming, especially for 3 interferers case where all three match in said alignments.

Our goal in the simulations is to obtain statistically significant BER performance results. For this, 3,000,000 navigation bits are simulated, which correspond to 1,000 minutes of GPS signals. We perform the simulation with a previously recorded signal file from the NI GPS simulator which is 200 seconds in duration. This recorded file has 12 visible GPS satellites at all times. We then proceed to run this file 300 times to obtain our navigation bits quota (10,000 bits x 300), while injecting interference bits randomly. This is all done internal with the proposed LabVIEW-based SDR.

## VI. CONCLUSION

This study shows strong interference mitigating capabilities with a proposed MMSE correlator solution, when compared to conventional matched filter correlators. The proposed SDR solution shows enhancement results as it works with real signals. An important value to this experiment is the capability of real-time operation. A scenario with 3 interferers, all synchronized in chip and bit alignment, which is considered a very low probability situation (less than 1%) showed gains of up to 7.3 dB when reduced complexity parameters were used such as $g=64$ and $l=1200$ for 3 interferers present, when comparing against conventional MF correlator.